\documentclass[twocolumn,amsmath,amssymb,aps,pre,superscriptaddress,longbibliography]{revtex4-1}

\usepackage{graphicx}
\usepackage{dcolumn}
\usepackage{bm}
\usepackage[english=american]{csquotes}
\usepackage[dvipsnames]{xcolor}

\begin{document}

\newcommand{\kb}{k_\text{B}}
\newcommand{\Av}[1]{\left\langle #1 \right\rangle}
\newcommand{\av}[1]{\langle #1 \rangle}
\newcommand{\n}{\nonumber}
\newcommand{\nn}{\nonumber \\}
\newcommand{\mb}{\bm}
\newcommand{\dif}[1]{\frac{\text{d}}{\text{d} #1}}
\newcommand{\ind}[1]{\int \text{d} #1}
\newcommand{\Ind}[1]{\int \text{d} \bm{#1}}

\title{Continuous-time random walk for a particle in a periodic potential}

\author{Andreas Dechant}
\affiliation{WPI-Advanced Institute of Materials Research (WPI-AIMR), Tohoku University, Sendai 980-8577, Japan}

\author{Farina Kindermann}
\affiliation{Department of Physics and Research Center OPTIMAS, University of Kaiserslautern, 67663 Germany}

\author{Artur Widera}
\affiliation{Department of Physics and Research Center OPTIMAS, University of Kaiserslautern, 67663 Germany}
\affiliation{Graduate School Materials Science in Mainz, 67663  Germany}

\author{Eric Lutz}
\affiliation{Institute for Theoretical Physics I, University of Stuttgart, 70550 Germany}

\begin{abstract}
Continuous-time random walks offer powerful  coarse-grained descriptions of transport processes. 
We here microscopically derive such a model for a Brownian particle diffusing in a deep periodic potential. We determine both the waiting-time and the jump-length distributions  in terms of the parameters of the system, from which we analytically deduce the non-Gaussian characteristic function. We  apply this continuous-time random walk model to characterize the underdamped diffusion of single Cesium atoms in a one-dimensional optical lattice. We observe excellent agreement between  experimental and  theoretical characteristic functions, without any free parameter.
\end{abstract}

\pacs{}

\maketitle

Coarse graining is an essential tool for the study of complex systems. Different levels of description of a system's  state are commonly identified \cite{esp04,cas08}. The microscopic (fine-grained) level corresponds to the true mechanical state. Its complete characterization is often out of reach, both  theoretically and experimentally, owing to its complexity.   By contrast,  the macroscopic (coarse-grained) regime consists of a few pertinent and accessible  variables that capture the main features of the system. All irrelevant degrees  of freedom are eliminated during  coarse-graining, for instance, by averaging over them. These various representations are usually associated with different time scales. Each of those may be used to define intermediate (mesoscopic) levels of coarse graining \cite{esp04,cas08}. Equilibrium thermodynamics is a prominent example of a coarse-grained theory where averaged quantities such as volume, pressure and temperature, are introduced to  specify the macroscopic state \cite{ehr56,jan63}. On the other hand, far from thermal equilibrium, the coarse-graining method is more involved as the  time evolution of the system needs to be accounted for \cite{esp04,cas08,ehr56,jan63,zwa01}.   Connecting fine-grained and coarse-grained descriptions  is in general  very challenging  for any realistic nonequilibrium system \cite{esp04,cas08,ehr56,jan63,zwa01}.

Continuous-time random walks (CTRW) are a well established approach  used to investigate the  coarse-grained nonequilibrium dynamics of complex systems \cite{hug96,kla11}. While  jumps occur at fixed discrete times in standard random walks, they happen at random continuous times in a CTRW. Continuous-time random walks provide general models for normal as well as anomalous diffusion and transport processes. They are completely characterized by a jump-length distribution and a waiting-time distribution between two jumps. They offer an effective method to compute the probability distribution of the position of the random walker from which various transport properties, such as moments and correlation functions, may be determined. Continuous-time random walks have been successfully applied in a wide range of areas, ranging from  disordered systems, plasmas and chaotic dynamics to turbulence, biology and finance \cite{han87,bou90,met00,hof13,zab15}. Experimental evidence for CTRW has been found for microbead motion in reconstituted actin networks \cite{won04}, lipid granules in cellular cytoplasm \cite{jeo11}, protein channels in plasma membranes \cite{wei11} and intermittent quantum dots \cite{sad14}. A few microscopic derivations of CTRW have been presented \cite{kla80,kes12}. However, in most cases, continuous-time random walks are  phenomenological as the complex microscopic dynamics is unknown. 

In this paper, we theoretically derive a CTRW model  for a Brownian particle in a deep periodic potential starting from the microscopic level and directly compare its predictions to experiment. This system is ubiquitous in physics, chemistry and biology \cite{ris89,cof04}. The microscopic description is based on the Langevin equation, a stochastic extension of Newton's equation of motion \cite{ris89,cof04}. This dynamics is not exactly solvable owing to the nonlinearity of the potential. The coarse-grained evolution is diffusive on very long  time scales with a Gaussian probability distribution. Diffusion coefficients are known in terms of the microscopic parameters both in the underdamped and overdamped regimes \cite{ris89,cof04}. However, no analytical results are available for the probability density or the position correlation function at intermediate  finite times, despite their experimental importance. A mesoscopic description is therefore needed. Such model should be simpler than the microscopic Langevin equation to allow an analytical description of the statistics of the process and, at the same time, more detailed than the macroscopic Gaussian diffusion approximation. 

In the following, we express the jump-length  and the waiting-time distributions in terms of microscopic variables of the system. From these two  distributions, we determine the Fourier-Laplace transform of the probability density, also known as the dynamical structure factor \cite{han87,bou90,met00,hof13,zab15}, as well as the characteristic function  with the help of the Montroll-Weiss formula for uncorrelated CTRW \cite{mon65}. We use the characteristic function to derive explicit expressions for the lower moments and the position correlation function of the  particle. We furthermore apply our theoretical results to describe the  underdamped diffusion of single Doppler-cooled Cesium atoms in a one-dimensional optical lattice \cite{gry01,met99,lut13}. We obtain very good agreement between experiment and theory for the characteristic function, both as a function of time and  wave vector, without any free parameter. We  finally determine the range of validity of the coarse-grained CTRW model from the experimental data and observe the transition to the macroscopic Gaussian diffusion approximation.

\textit{Continuous-time random walk model.}
The continuous-time random walk  extends the concept of a random walker, which randomly takes unit steps to the left or to the right at discrete time intervals, to a continuous process with arbitrary step sizes.
The  main ingredients of the CTRW are the jump-length distribution $\phi(\xi)$  and the waiting-time distribution $\psi(\tau)$. When these distributions are mutually independent and the same for each step, they define a renewal process \cite{hug96,kla11}. In that case, the probability distribution $P(x,t)$ is simply related to the waiting-time and jump-length distributions in Fourier-Laplace space through the Montroll-Weiss formula \cite{mon65},
\begin{align}
S(k,s) = \frac{1 - \tilde{\psi}(s)}{s \big[1 - \hat{\phi}(k) \tilde{\psi}(s) \big]} \label{montroll-weiss},
\end{align}
The dynamical structure factor $S(k,s)$ is here the Fourier-Laplace transform of  $P(x,t)$, and $\hat{\phi}(k)$ and $\tilde{\psi}(s)$ are the respective Fourier and Laplace transforms of the jump-length and waiting-time distributions. 
Equation \eqref{montroll-weiss} fully characterizes the statistics of the CTRW at arbitrary times for given distributions $\phi(\xi)$ and  $\psi(\tau)$. 
The latter are often determined phenomenologically. 
We will next explicitly derive both distributions for a  Brownian particle moving in a periodic potential.

Let us consider a classical particle of mass $m$ in contact with a heat bath at temperature $T$ moving in a periodic potential, $U(x+L) = U(x)$, of period $L$ and depth $U_0$. 
Its microscopic dynamics is governed  by the underdamped Langevin equation for the velocity $v(t)$ \cite{ris89,cof04},
\begin{align}
\dot{v}(t) = - \gamma v(t) - \frac{1}{m} U'(x(t)) + \frac{\sqrt{2 \gamma \kb T}}{m} \eta(t) , \label{langevin}
\end{align}
where $\eta(t)$ denotes a centered and delta-correlated Gaussian white noise, $\av{\eta(t) \eta(t')} = \delta(t-t')$, $\gamma$ is the damping coefficient and  $\kb$  the Boltzmann constant.
In order to cast the dynamics described by Eq.~\eqref{langevin} in terms of a CTRW, we need to coarse-grain and decompose it into a series of consecutive jump and waiting events.
This can be done for deep potentials, $U_0 \gtrsim 4\kb T$, where the particle spends most of the time close to one of the minima of the potential and only occasionally escapes to another well \cite{fer93}.
We identify these trapping periods with the waiting times and the escape events with the jumps of the CTRW.
In this limit of well separated time scales, the escape process  may be described by Kramers' rate theory, which states that the fraction of initially trapped particles remaining in a potential well decays over time at a rate $1/\tau_0$ \cite{han90}.
This immediately translates into an exponential distribution for the escape times,
\begin{align}
\psi(\tau) = \frac{1}{\tau_0} e^{-\frac{\tau}{\tau_0}} \label{waiting-time-dist}.
\end{align}
Performing the Laplace inversion of the dynamical structure factor \eqref{montroll-weiss}, we thus obtain the characteristic function $K(k,t)$ in the time-domain,
\begin{align}
K(k,t) &= e^{-\frac{t}{\tau_0} \big[1 - \hat{\phi}(k)\big]} \label{CTRW-model} .
\end{align}
Compared to the probability distribution $P(x,t)$, this representation has two advantages:
First, the dependence on the wave vector $k$ for fixed time is simply related to the Fourier transform $\hat{\phi}(k)$ of the jump-length distribution.
Second, at fixed $k$, the decay of the characteristic function is explicitly exponential in time, allowing  to easily verify the validity of Eq.~\eqref{waiting-time-dist} for a given jump process.

For a standard cosine-shaped potential, $U(x) = U_0[1 - \cos(2 \pi x/L)]/2$, the escape time $\tau_0$ can be estimated  both in the strong damping and  weak damping limits \cite{han90},
\begin{eqnarray}
\tau_0 \simeq \left\lbrace \begin{array}{ll}
\frac{\pi \gamma}{\omega_0^2} e^{\frac{U_0}{\kb T}} &\text{for} \quad \gamma \gg \omega_0, \\[2 ex]
\frac{\pi}{4 \gamma} \frac{\kb T}{U_0} e^{\frac{U_0}{\kb T}} &\text{for} \quad \gamma \ll \omega_0  ,
\end{array} \right. \label{escape-time}
\end{eqnarray}
where $\omega_0 = 2 \pi \sqrt{2 U_0/(m L^2)}$ is the curvature of the potential at the bottom of the well.
Both expressions are valid in the limit of deep potentials $U_0 \gg \kb T$.
This timescale is exponential in the potential depth and increases as either the high- or low-dissipation limit is approached.
We note that more involved formulas that provide a better approximation for moderately deep potentials may also be obtained (see Appendix \ref{app-over-under}).

 For strong damping, the particle immediately equilibrates after escaping to a neighboring well and thus the probability of jumping multiple lattice sites is negligible \cite{fer93}.
In this limit,  the jump length distribution is simply $\phi(\xi) = \delta(\xi \pm L)/2$.
By contrast, in the weak damping regime, once the particle attains an energy sufficient for escaping, the relaxation of the energy back towards the thermal average is slow, and the particle can jump over multiple wells before becoming trapped again.
An expression for the jump-length distribution, valid at low dissipation, may be obtained from the discrete probability $\phi^*(n)$ of jumping $n$ lattice sites of the periodic potential in either direction derived by Mel'nikov \cite{mel91},
\begin{align}
\phi^*(n) &= N \mathcal{P}\bigg(\frac{\mathcal{E}}{\kb T} n \bigg) \label{jump-dist-under} \\
\text{with} \quad \mathcal{P}(y) &= e^{-\frac{y}{4}} \int_0^\infty \text{d}z \ \frac{z^2 e^{- \frac{y z^2}{4}}}{\big(1 + \sqrt{2 + z^2}\big)^2} \n ,
\end{align}
where $N$ is a normalization constant such that $\sum_{n=1}^\infty \phi^*(n) = 1$. The continuous jump-length distribution follows as $\phi(\xi) = \phi^*(n) \delta(|\xi| - n L)$. The quantity $\mathcal{E}$ in Eq.~\eqref{jump-dist-under} denotes the energy dissipated by a particle traveling a distance $L$ at an energy $U_0$ that is just sufficient to escape from a well. It is  given by  \cite{mel91},
\begin{align}
\mathcal{E} = \sqrt{m} \gamma \int_0^L \text{d}x \ \sqrt{2 [U_0 - U(x)]} \label{energy-diss}.
\end{align}
For the cosine-shaped potential, the integral \eqref{energy-diss} can be evaluated explicitly and yields $\mathcal{E} = 2 \gamma L \sqrt{2 m U_0}/\pi$.
It is important to note that the function $\mathcal{P}(y)$ behaves as $\mathcal{P}(y) \propto y^{-1/2}$ for small $y$ and as $\mathcal{P}(y) \propto y^{-3/2} e^{-y/4}$ for large $y$. As a consequence,  the tails of the jump-length distribution are not exactly exponential.

Because of the exponential waiting-time distribution \eqref{waiting-time-dist}, the CTRW  is memoryless.  The $n$-point  probability distributions hence factorize \cite{hug96,kla11}. The $2$-point probability density can, for example, be written as the product,
\begin{align}
P(x_2,t_2;x_1,t_1) = P(x_2-x_1,t_2-t_1) P(x_1,t_1) \label{poisson-factorize} .
\end{align}
 We can accordingly express arbitrary $n$-point correlation functions in terms of the characteristic function $K(k,t)$ using Eq.~\eqref{CTRW-model} (see Appendix \ref{app-corr}). The position $2$-point correlation function is, for instance, given by,
\begin{eqnarray}
\Av{x(t_2) x(t_1)} &=& - \partial_k \Big[ K(k,t_2-t_1) \partial_k K(k,t_1) \Big] \bigg\vert_{k = 0},\nonumber \\
&=& - \frac{t_1}{\tau}\, \partial_k^2 \hat \phi(k)|_{k = 0} = \frac{t_1}{\tau} \,\langle\xi^2\rangle.\label{corr}
\end{eqnarray}
In addition, the second and fourth moments read, 
\begin{eqnarray}
\av{\Delta x^2(t)} &=& \frac{t}{\tau_0}\av{\xi^2}, \label{10}  \\ \av{\Delta x^4(t)} &=& \frac{t^2}{\tau_0^2} \big( 3 \av{\xi^2}^2 + \frac{\tau_0}{t} \av{\xi^4} \big) \label{moments} ,
\end{eqnarray}
where we have defined  the displacement $\Delta x(t) = x(t) - x(0)$. The second moment \eqref{10} is linear in time indicating that  the position of the particle  exhibits normal diffusion with diffusion coefficient $D_x = \av{\xi^2}/(2\tau_0)$ at all times, as in the asymptotic  Gaussian diffusion limit \cite{ris89,cof04}. By contrast, the displacement distribution  is not Gaussian at finite times. The departure from Gaussianity may be quantified with the excess kurtosis \cite{ris89,cof04},
\begin{align}
\kappa(t) = \frac{\av{\Delta x^4(t)}}{3 \av{\Delta x^2(t)}^2} - 1 = \frac{\av{\xi^4}}{3 \av{\xi^2}^2} \frac{\tau_0}{t}  \label{ngp} .
\end{align}
This quantity is zero for Gaussian distributions. 
It is positive for the CTRW indicating that large displacements are more prevalent than in the Brownian case. 
The Gaussian limit is recovered for large times as $1/t$ with a rate that is controlled by the excess kurtosis of the step-size distribution, $\kappa_\xi={\av{\xi^4}}/({3 \av{\xi^2}^2})$. 
Due to the algebraic decay, deviations from Gaussian diffusion can be significant even at longs times.

\textit{Single atoms in an optical lattice.} 
Our CTRW model \eqref{CTRW-model} holds for any deep periodic potential, both in the underdamped and overdamped limits (see Appendix \ref{app-over-under}).
In order to test its predictions  and determine its range of validity, we now apply it to experimental data obtained by measuring the motion of single atoms in an optical lattice \cite{kin17}.
In the experiment, a Cesium atom is trapped in the periodic potential of a one-dimensional optical lattice with $U_0/\kb \approx 210 \ \mu \text{K}$ and $L = \lambda/2$ where $\lambda = 790 \ \text{nm}$ is the wavelength of the lattice beam. 
Damping at a rate of $\gamma \approx 5 \cdot 10^3 \ \text{s}^{-1}$ is due to a Doppler cooling force and noise  is induced by random absorption and emission of photons, resulting in a recoil of the atom \cite{met99}.
Both  provide an effective thermal bath  at a temperature of $T \approx 50 \ \mu \text{K}$. 
The thermal energy  is more than four times smaller than the  lattice depth, corresponding to the deep-potential limit. 
Quantum tunneling is suppressed by frequent photon scattering with rates in the MHz regime. The atomic motion can thus be treated classically.

The jump-length distribution \eqref{jump-dist-under} is  entirely specified by the energy dissipation per period of the lattice \eqref{energy-diss} which is equal to $\mathcal{E} \approx 0.13 \kb T$ in the experiment. 
This value  corresponds to the weak damping regime $\gamma \ll \omega_0$.
 On the other hand, the  waiting-time distribution \eqref{waiting-time-dist} is fully characterized by the escape time $\tau_0$ from a potential well, which for the experimental parameters is given  as $\tau_0 \approx 3.4 \ \text{ms}$ (see Appendix \ref{app-over-under}). 
 This is the central relevant time scale for the motion of the atom in the optical lattice.
 
\begin{figure}[t]
\includegraphics[width=.47\textwidth]{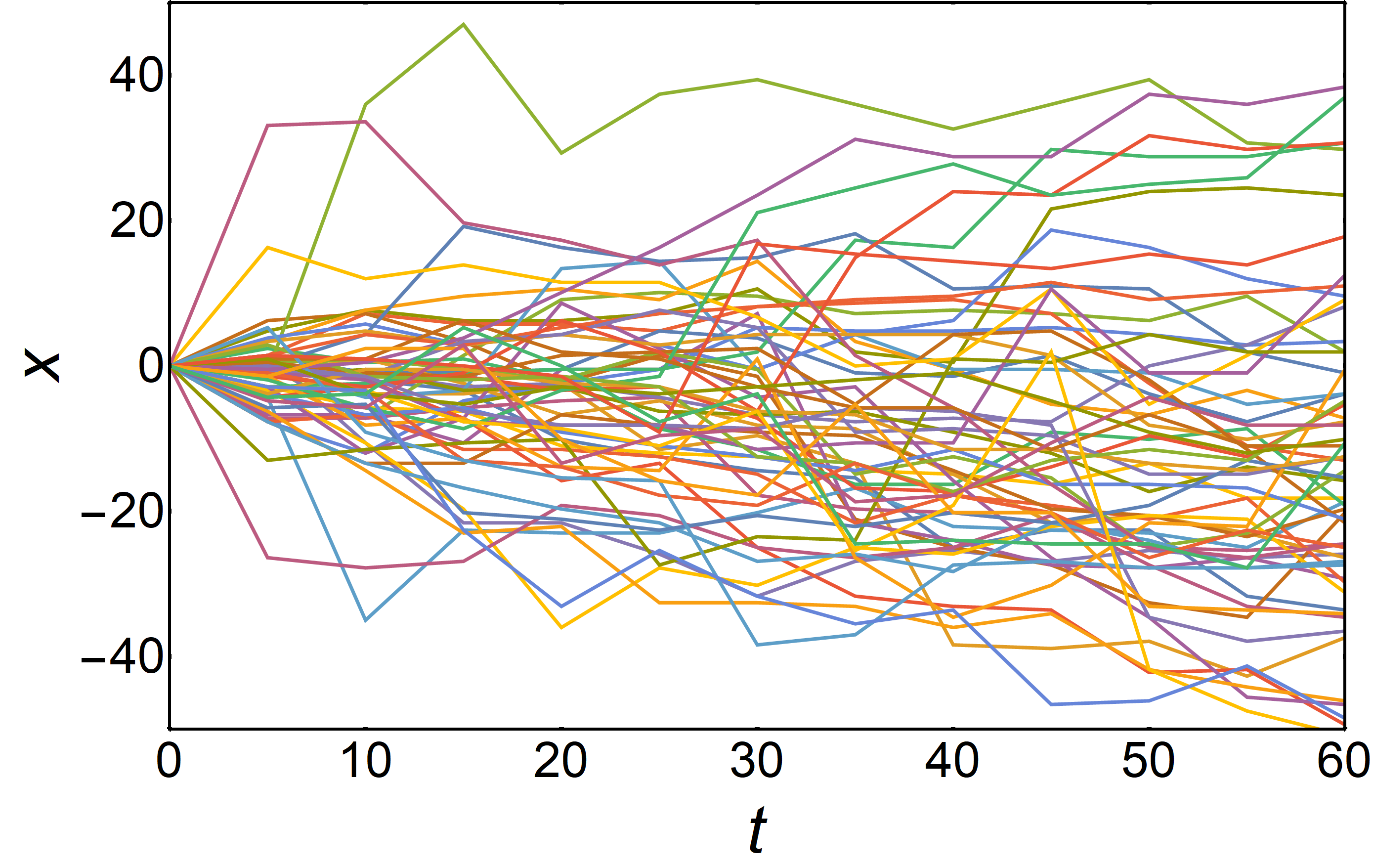}
\caption{Position of single Cesium atoms diffusing in a one-dimensional optical lattice  as a function of  time. The various lines show 60 randomly chosen coarse-grained traces for a time between stroboscopic position measurements of $\tau_\text{flight} = 5$ ms. }
\end{figure}

\begin{figure*}[t]
\includegraphics[height=4.2cm]{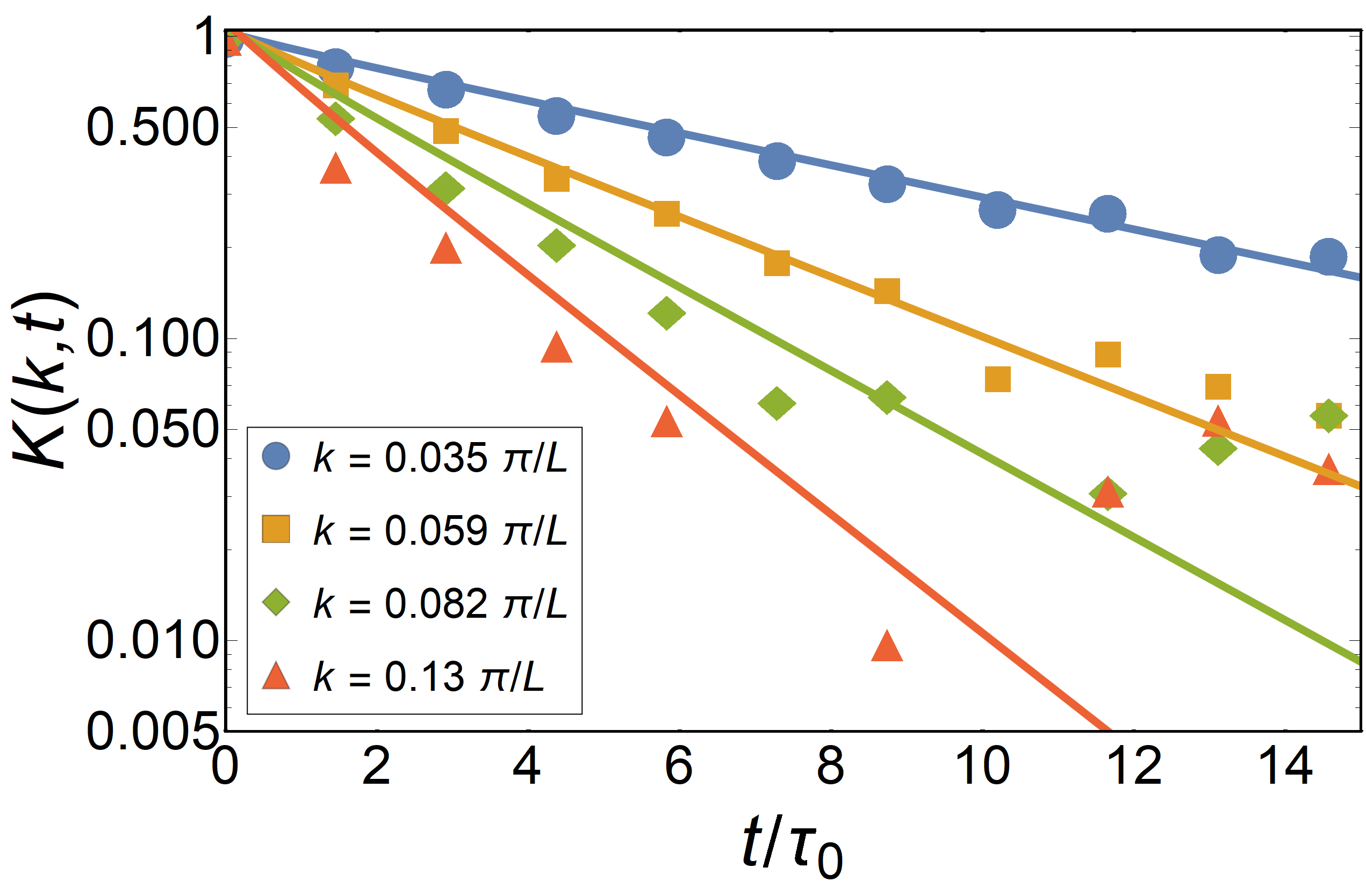}
\includegraphics[height=4.2cm]{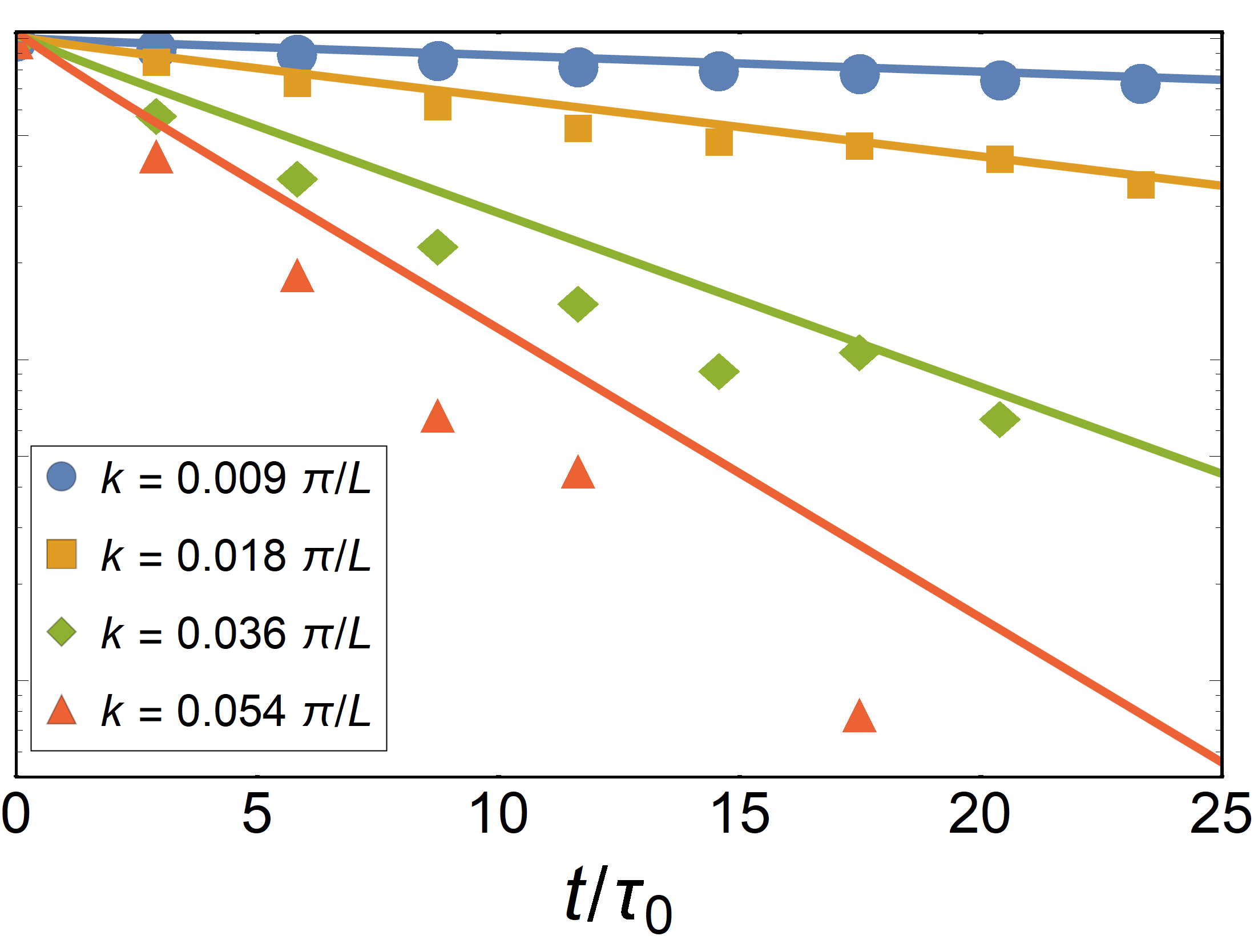}
\includegraphics[height=4.2cm]{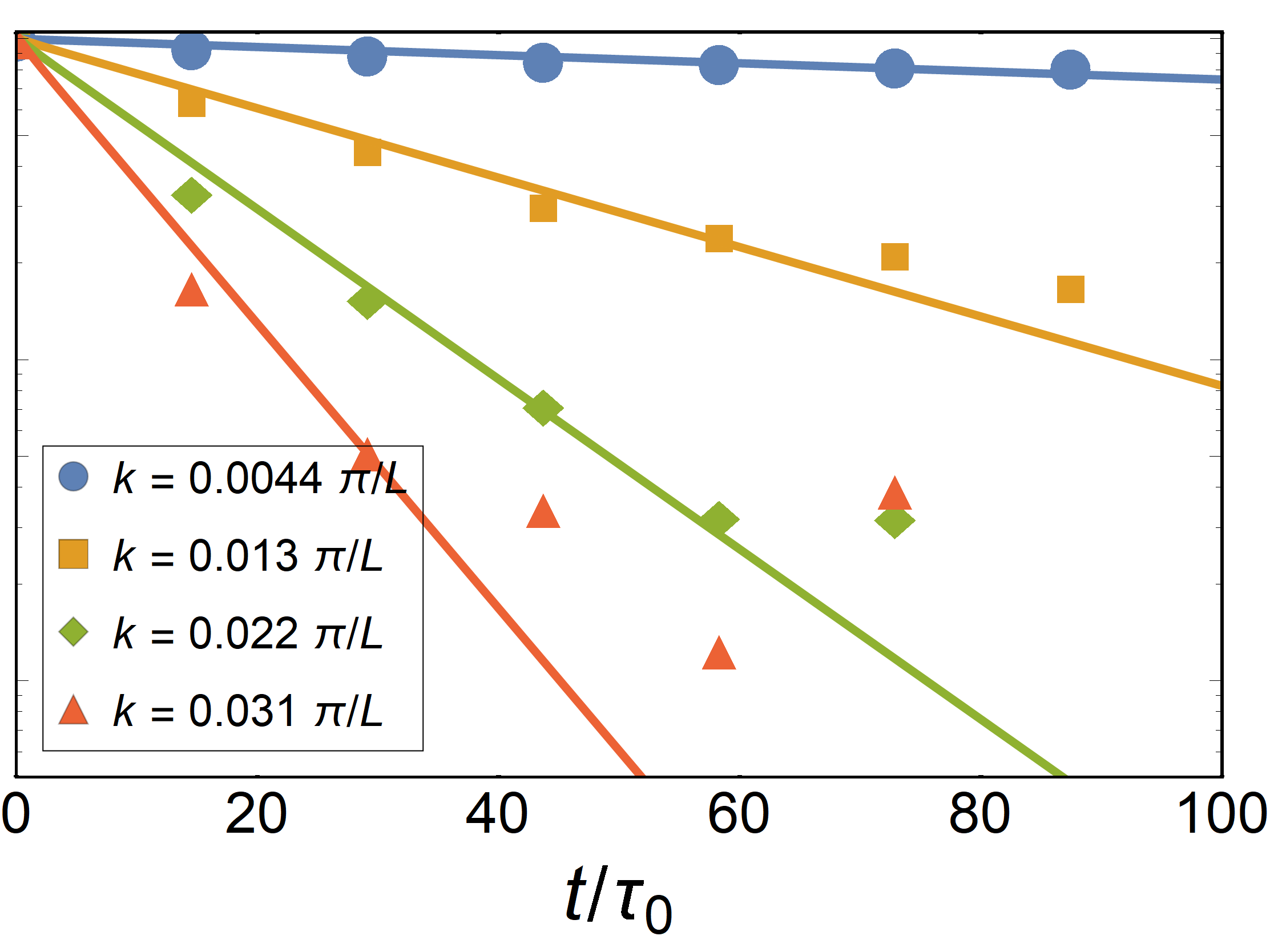}\\
\includegraphics[height=4.1cm]{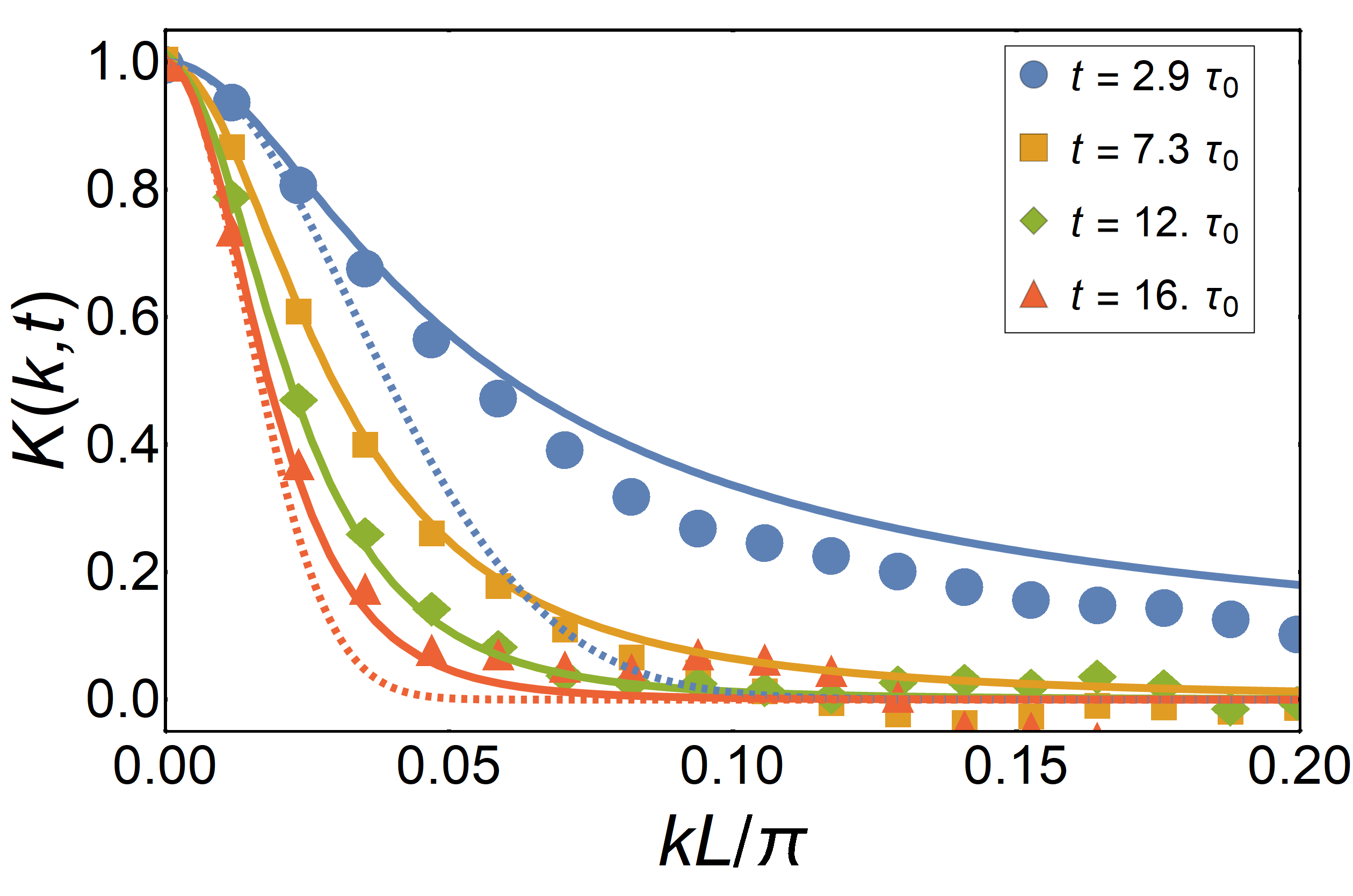}
\includegraphics[height=4.1cm]{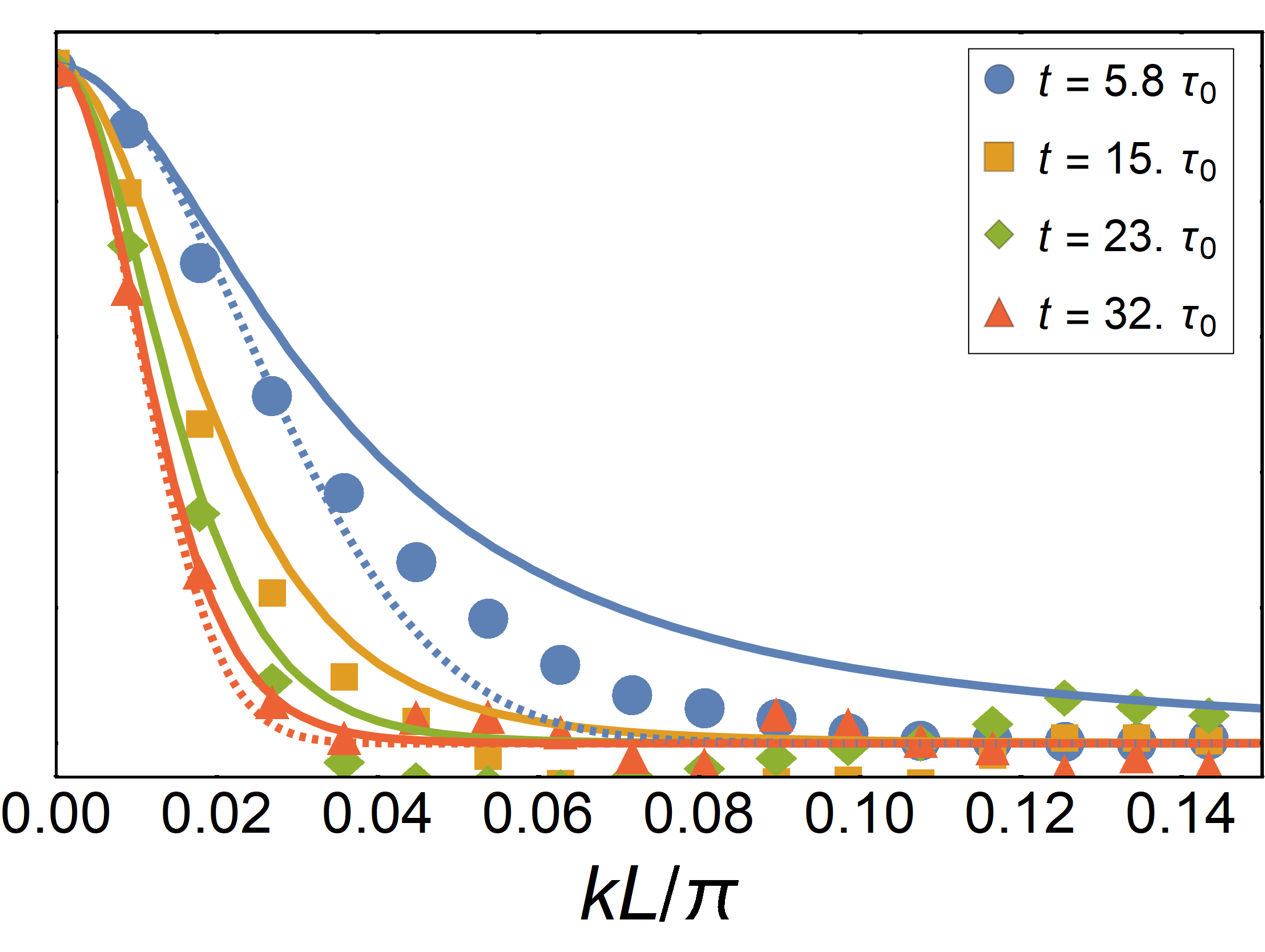}
\includegraphics[height=4.1cm]{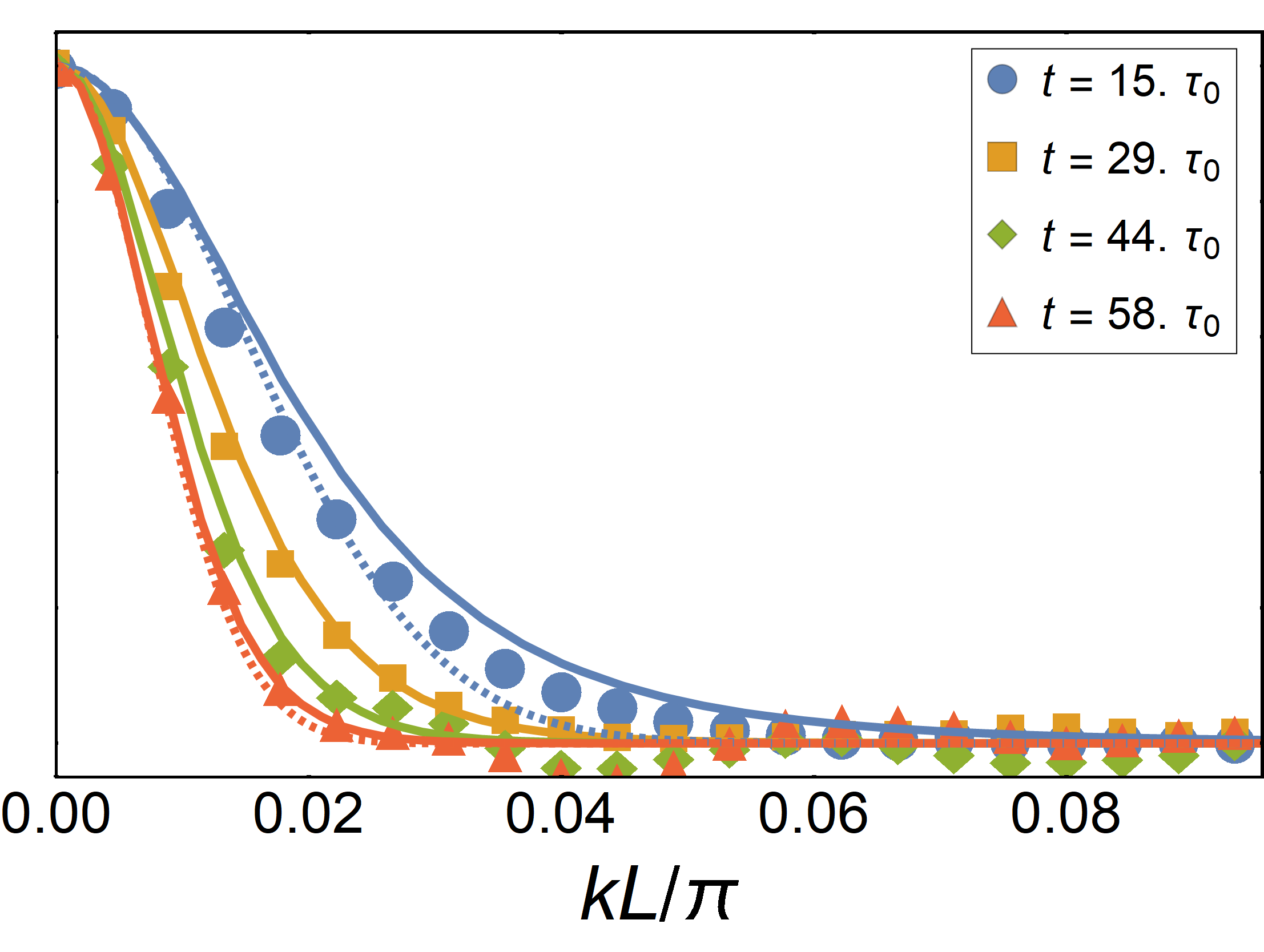}
\caption{Characteristic function $K(k,t)$  of single Cesium atoms diffusing in a one-dimensional optical lattice as a function of time $t$ (top) and wave vector $k$ (bottom).
The symbols correspond to the experimental data, the solid lines to the analytical prediction Eq.~\eqref{CTRW-model} with Eq.~\eqref{jump-dist-under}, evaluated using the experimental parameters $\mathcal{E} = 0.13 \kb T$ and $\tau_0 = 3.4 \ \text{ms}$, without any free parameter.
The first, second and third column correspond to  $\tau_\text{flight} = 5 \ \text{ms}, \ 10 \ \text{ms}, \ 50 \ \text{ms}$, respectively.
We observe very good agreement between  data and theoretical  predictions, in particular for small values of $k$, which reflect the long-range behavior of the jump process.
From the panels in the top row, we see that the time-dependence of the characteristic function at fixed $k$ is well-described by an exponential decay, confirming the validity of the exponential waiting time distribution \eqref{waiting-time-dist}.
On the other hand, the panels in the bottom row show that also the predicted dependence on $k$ resulting from the jump length distribution \eqref{jump-dist-under} is well-reproduced in the experiment. 
It further provides a much better description of the data than the asymptotic  Gaussian diffusion approximation with  the same diffusion coefficient (dashed lines), in particular at short times (blue). \label{fig-charfunc-exp}}
\end{figure*}

The position of the atom  after a time $\tau_\text{flight}$ is measured by ramping up the potential to $U^* \approx 850 \ \mu \text{K}$ while the cooling beam is switched off, effectively immobilizing the atom, and  taking a high-resolution fluorescence image.
Subsequently, the potential is lowered to $U_0$, allowing the atom to move again for a time $\tau_\text{flight}$.
Repeating this procedure 14 times for every atom generates a coarse-grained measurement of the diffusion process. 
For each parameter set, 600 to 1000 atomic trajectories are recorded (typical examples are shown in Fig.~1). 
The longest traces spread over approximately 100 lattice sites corresponding to a total distance of 40$\mu$m. 
The position resolution is about 2$\mu$m. 
A high number of photons (of the order of $10^6$) are scattered during the image taking. 
As a consequence, the particle loses its memory about previous steps and jumps are independent of each other.

Figure \ref{fig-charfunc-exp} displays the characteristic function $K(k,t)$, both as a function of time $t$ (top) and of the wave vector $k$ (bottom) for increasing values of $\tau_\text{flight} = 5 \ \text{ms}$, $ 10 \ \text{ms}$ and $ 50 \ \text{ms}$. 
The symbols correspond to the experimental data and the lines to the analytical predictions given by Eq.~\eqref{CTRW-model} together with the jump-length distribution based on Eq.~\eqref{jump-dist-under}, evaluated using the experimental parameters $\mathcal{E} = 0.13 \kb T$ and $\tau_0 = 3.4 \ \text{ms}$. 
We observe overall remarkable agreement between theory and experiment, without any free parameter. 
In particular, the $t$-dependence of the characteristic function (left) confirms the exponential form of the waiting-time distribution \eqref{waiting-time-dist}, while the $k$-dependence (right) corroborates the step-size distribution \eqref{jump-dist-under}. 
Deviations are seen for large values of $k$, that is, at short distances. These are, on the one hand, due to the finite spatial resolution of the imaging process. On the other hand, we note that, since the motion of the atoms is only weakly damped, the atoms do not immediately thermalize after a jump. They thus have an increased probability to jump again.
These repeated-jump events are not captured by the CTRW model. They cause the characteristic function to depart from Eq.~\eqref{CTRW-model} for values of $k L/\pi \gtrsim 0.07$, corresponding to length scales of a few individual jumps (see Appendix \ref{app-num} for a more detailed discussion). Such deviations do not occur in the overdamped regime. 
Comparing the CTRW of the atoms with standard Gaussian diffusion, the difference is clearly visible at short times  (blue dashed lines in Fig.~\ref{fig-charfunc-exp}, in particular for the short flight time $\tau_\text{flight} = 5$ ms).
On time scales that are very long compared with the escape time $\tau_0$, on the other hand, the position of the particle is the sum over many independent jump events, and the distribution converges towards a Gaussian in accordance with the central-limit theorem (red dashed lines).

\textit{Conclusions.} We have derived a  continuous-time random walk model for  a Brownian particle in a periodic potential.
This model represents an intermediate level of coarse-graining between the full microscopic dynamics and a simple diffusion approximation.
As such, it permits a detailed, yet analytical characterization of the statistics of the process. It is valid for deep potentials, $U_0\gtrsim 4k_BT$, both in the underdamped and overdamped regimes.
We have concretely determined the waiting-time and jump-length distributions, from which we have obtained the dynamical structure factor and the non-Gaussian characteristic function. 
We have, additionally, observed excellent agreement between theoretical predictions and experimental data for the weakly damped diffusion of single laser-cooled Cesium atoms moving in a one-dimensional optical lattice, without any free parameter. 
Our results establish a transparent and useful bridge between microscopic and macroscopic theoretical descriptions of a paradigmatic nonequilibrium system and, at the same time, between analytical formulas and experiment.

\begin{acknowledgments}
\textit{Acknowledgments.} This work was supported by the World Premier International Research Center Initiative (WPI), MEXT, Japan, and by Deutsche Forschungsgemeinschaft (DFG) via Sonderforschungsbereich (SFB) 49.
\end{acknowledgments}

\appendix

\begin{widetext}
\renewcommand\thesection{\Alph{section}}
\renewcommand{\theequation}{\thesection\arabic{equation}}

\section{Overdamped and underdamped approximations} \label{app-over-under}

\setcounter{equation}{0}
We here discuss the overdamped and underdamped approximations of the decay time $\tau_0$. While the solution of the Langevin equation \eqref{langevin}
involving a periodic potential,
\begin{align}
U(x) = \frac{U_0}{2} \Big[1 - \cos\Big(\frac{2 \pi x}{L} \Big) \Big], \label{potential}
\end{align}
is not tractable in an analytic manner in full generality, there are two limits that are of great practical importance, both of which rely on a separation of time scales \cite{ris89}.
In the overdamped limit $\gamma \gg \omega_0$, where $\omega_0 = 2 \pi \sqrt{2 U_0/(m L^2)}$ is the curvature  at the bottom of a well, the particle is strongly coupled to the heat bath and the relaxation of its velocity degree of freedom is much faster than the motion of the particle in the potential.
In this limit, we can neglect the inertial term in Eq.~\eqref{langevin} to obtain the overdamped Langevin equation for the slow position degree of freedom \cite{ris89},
\begin{align}
m \gamma \dot{x}(t) = -  U'[x(t)] + \sqrt{2 \gamma \kb T} \xi(t), \label{overdamped}
\end{align}
or, equivalently, the Smoluchowski equation for the probability density $P_x(x,t)$,
\begin{align}
\partial_t P_x(x,t) = \frac{1}{m \gamma} \partial_x \Big[ U'(x) + \kb T \partial_x \Big] P_x(x,t) .
\end{align}
Let us assume that the particle is initially located at the bottom of the potential well at $x=0$.
The particle can escape from the potential well to a neighboring well, if, due to the random kicks received from particles of the surrounding medium, it diffuses to $x = \pm L$.
Note that since the particle is heavily damped, the probability of thermalizing in the neighboring well is close to unity and we can neglect jumps over multiple wells at the same time.
This results in a jump length distribution of $\phi(\xi) = \delta(\xi \pm L)/2$.
The escape time is given by the average time it takes the particle to first reach the potential hill at $x = \pm L/2$, starting from $x = 0$, the so-called first passage time.
The average first passage time $\tau(x)$ starting from a point $x$ follows the differential equation \cite{ris89},
\begin{align}
\frac{1}{m \gamma} \Big[ U'(x) \partial_x + \kb T \partial_x^2 \Big] \tau(x) = -1 .
\end{align}
with boundary conditions $\tau(\pm L/2) = 0$.
The solution is given by \cite{han90}
\begin{align}
\tau(x) = \frac{m \gamma}{\kb T} \int_{x}^\frac{L}{2} dy \int_{-\frac{L}{2}}^y dz \ e^{\frac{U(y)-U(z)}{\kb T}} \label{fpt-over}.
\end{align}
For $x = 0$ and in the limit of very deep potentials $U_0 \gg \kb T$, this reduces to
\begin{align}
\tau_0 \simeq \frac{m \gamma L^2}{8 \pi U_0} e^{\frac{U_0}{\kb T}} \label{fpt-over-deep},
\end{align}
which is the first line of Eq.~\eqref{escape-time}.

In the opposite limit of very weak coupling between particle and heat bath $\gamma \ll \omega_0$, depending on the total energy $E = m v^2/2 + U(x)$, the particle either oscillates in a potential well (for $E < U_0$) or moves freely through the lattice (for $E > U_0$).
Importantly, the energy of the particle is almost conserved during the time it takes to complete one oscillation or move a distance $L$ trough the lattice. 
In this case, the energy becomes the slow degree of freedom, and its probability distribution $P_E(E,t)$ is governed by the Fokker-Planck equation \cite{ris89},
\begin{align}
\partial_t P_\varepsilon(\varepsilon,t) = 2 \gamma \partial_\varepsilon \bigg[ g(\varepsilon) + \frac{\kb T}{U_0} g(\varepsilon) \partial_\varepsilon \bigg] \frac{P_\varepsilon(\varepsilon,t)}{h(\varepsilon)} \label{energy-fpe} .
\end{align}
Here $\varepsilon = E/U_0$ is the energy in units of the potential barrier and we have defined the functions
\begin{align}
h(\varepsilon) = \left\lbrace \begin{array}{ll}
\mathrm{K}(\varepsilon) &\text{for} \quad \varepsilon < 1 \\[1ex]
\frac{1}{\sqrt{\varepsilon}} \mathrm{K}\Big(\frac{1}{\varepsilon}\Big) &\text{for} \quad \varepsilon > 1
\end{array} \right. \qquad \text{and} \qquad g(\varepsilon) = \left\lbrace \begin{array}{ll}
\sqrt{\varepsilon}\mathrm{E}\Big(\arcsin(\sqrt{\varepsilon}),\frac{1}{\varepsilon}\Big) &\text{for} \quad \varepsilon < 1 \\[1ex]
\sqrt{\varepsilon} \mathrm{E}\Big(\frac{1}{\varepsilon}\Big) &\text{for} \quad \varepsilon > 1 .
\end{array} \right.
\end{align}
Here $\mathrm{K}(x)$ is the complete elliptic integral of the first kind and $\mathrm{E}(x)$ ($\mathrm{E}(\phi,x)$) the complete (incomplete) elliptic integral of the second kind \cite{whi90}.
Physically, these functions represent the time $\tau_\text{osc} = 4 h(\epsilon)/\omega$ required for a particle of energy $\varepsilon U_0$ to complete one oscillation or move a distance $L$ and the change in energy $\Delta E = 4 U_0 \gamma g(\epsilon)/\omega$ during this time.
Note that $g(\varepsilon)$ is continuous at $\varepsilon = 1$, whereas $h(\varepsilon)$ exhibits a logarithmic divergence at this point, since a particle at exactly the barrier energy takes an infinite amount of time to complete one oscillation.
From this, we can compute the steady state energy distribution,
\begin{align}
P_{\varepsilon,\text{st}}(\varepsilon) = N h(\varepsilon) e^{-\epsilon \frac{U_0}{\kb T}} \label{energy-dist-stat},
\end{align}
where $N$ is a normalization constant such that $\int_0^\infty d\varepsilon \ P_{\varepsilon,\text{st}}(\varepsilon) = 1$.
Equation \eqref{energy-fpe} allows us to translate the problem of finding the average escape time into the calculation of the first passage time for the particle's energy to reach $\varepsilon = 1$, starting from $\varepsilon_0 < 1$, which is given by,
\begin{align}
\tau(\varepsilon_0) = \frac{U_0}{2 \gamma \kb T} \int_{\varepsilon_0}^1 dy \ \frac{e^{\frac{y U_0}{\kb T}}}{g(y)} \int_0^y dz \ h(z) e^{-\frac{z U_0}{\kb T}} \label{fpt-under}.
\end{align}
Similar to Eq.~\eqref{fpt-under}, this expression can be simplified for $\varepsilon_0 = 0$ and in the limit $U_0 \gg \kb T$ to yield,
\begin{align}
\tau_0 \simeq \frac{\pi \kb T}{4 \gamma U_0} e^{\frac{U_0}{\kb T}} \label{fpt-under-deep},
\end{align}
which is the second line of Eq.~\eqref{escape-time}.
However, while the simplified expression \eqref{fpt-over-deep} is numerically accurate in the overdamped limit even at moderately deep potentials (less than $10 \%$ deviation from Eq.~\eqref{fpt-over} for $U_0 > 5 \kb T$), the underdamped result converges to the asymptotic expression \eqref{fpt-under-deep} much more slowly (about $50 \%$ deviation from Eq.~\eqref{fpt-under} at $U_0 = 5 \kb T$) and therefore Eq.~\eqref{fpt-under} should be used.
Assuming that the initial energy of the particles is distributed according to Eq.~\eqref{energy-dist-stat}, we obtain for the typical escape time,
\begin{align}
\tau_0 = \frac{U_0}{2 \gamma \kb T \int_0^1 dx \ h(x) e^{-x \frac{U_0}{\kb T}}} \int_0^1 dx \int_x^1 dy \int_0^y dz \ e^{-(x+z-y)\frac{U_0}{\kb T}} \frac{h(x) h(z)}{g(y)} \label{fpt-under-average} .
\end{align}

\section{Correlation functions and characteristic function} \label{app-corr}
\begin{figure}[t]
\includegraphics[width=0.47\textwidth]{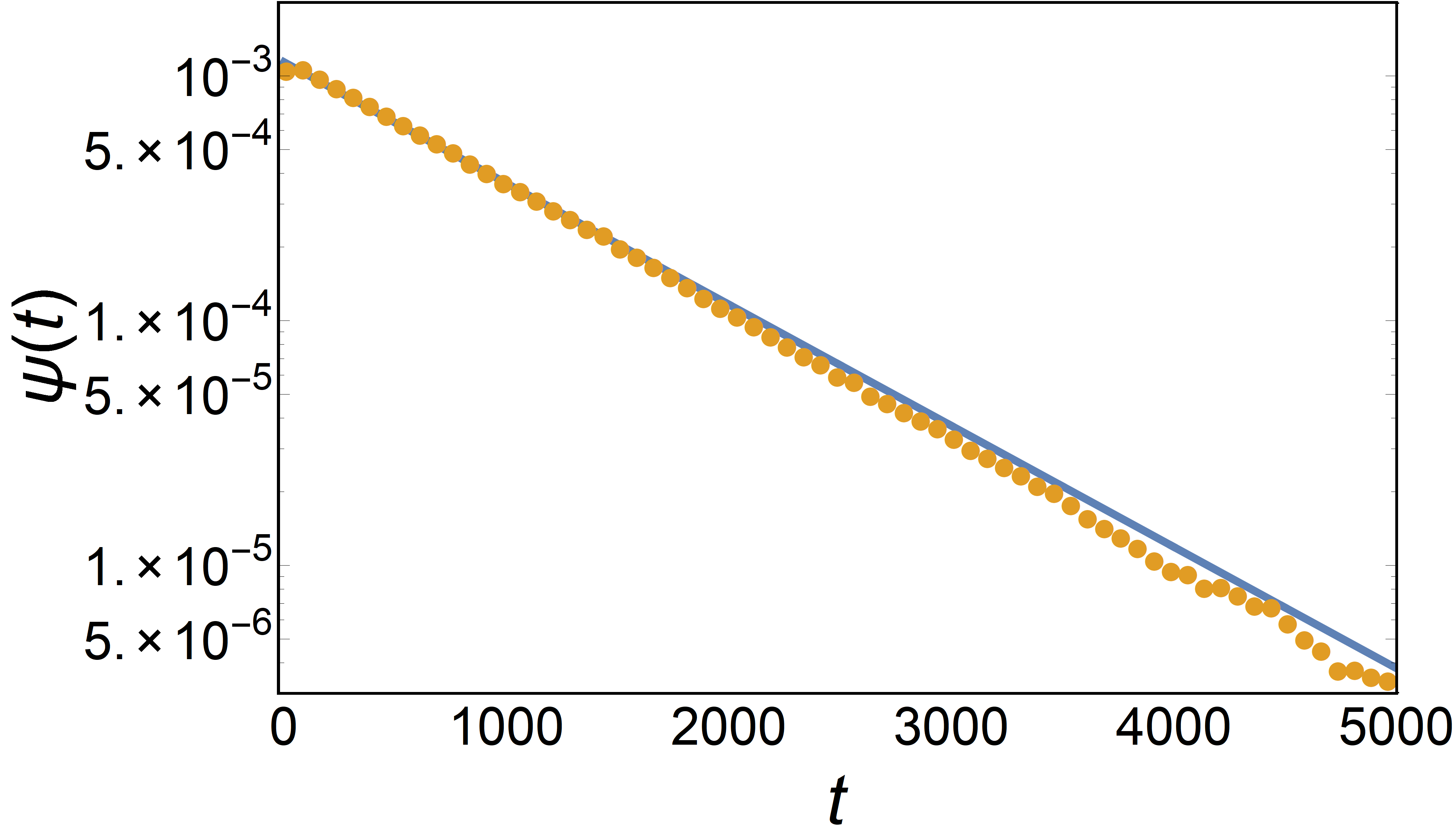}
\caption{Comparison of the numerically obtained waiting time distribution in the strong damping regime with the theoretical predictions.
The orange dots represent the numerical data, the blue lines are an exponential distribution with time constant $\tau_0$ given by Eq.~\eqref{fpt-over-deep}. The depth of the lattice is $U_0 = 5 \kb T$ and $m \gamma  = 1$.} \label{fig-esc-jump-over}
\end{figure}
\begin{figure}[t]
\includegraphics[width=0.47\textwidth]{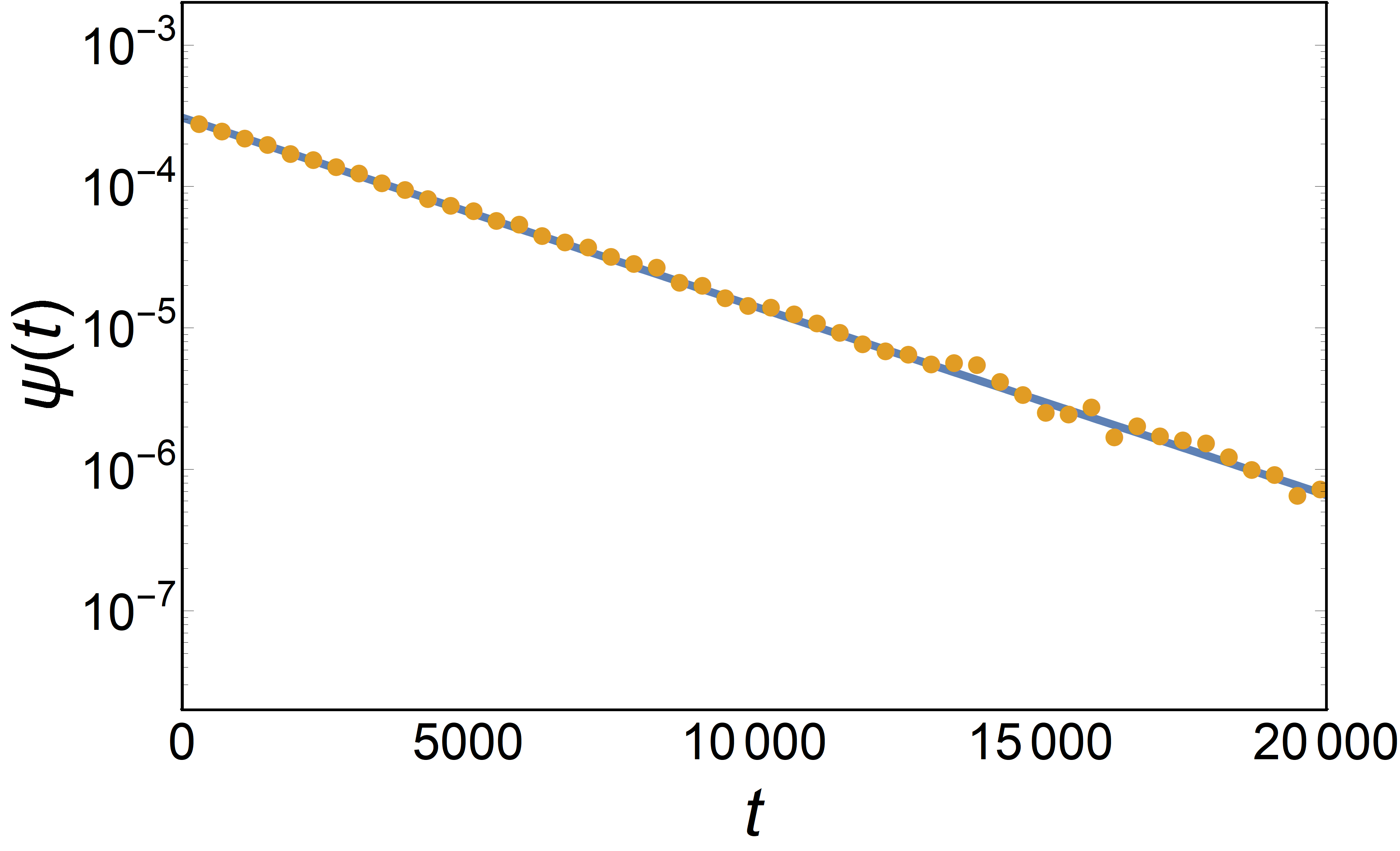}
\includegraphics[width=0.49\textwidth]{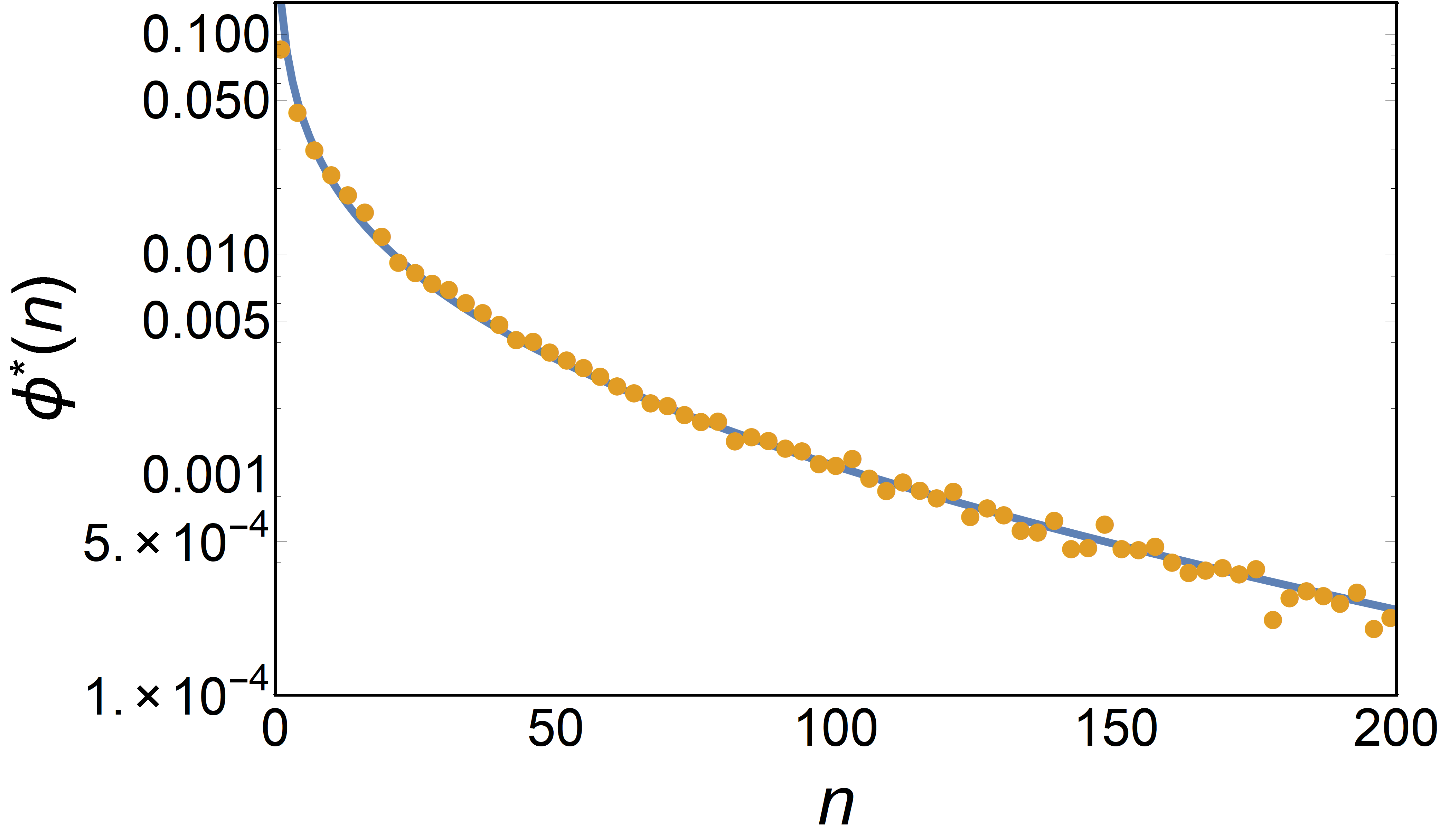}
\caption{Comparison of the numerically obtained waiting time (left) and jump length (right) distributions in the weak damping regime with the theoretical predictions.
The orange dots represent the numerical data, the blue lines are an exponential distribution with time constant $\tau_0$ given by Eq.~\eqref{fpt-under-average} (left) and the jump length distribution given in Eq.~\eqref{jump-dist-under}.
Here we use $U_0 = 5 \kb T$ and $\gamma=0.01$ (corresponding to $\mathcal{E} = 0.023 \ \kb T$)}, the remaining parameters  are $m = 1$, $U_0 = 1$ and $L = 0.5$. \label{fig-esc-jump-under}
\end{figure}
In Eq.~\eqref{corr}, the expression for the two-point correlation function in terms of the characteristic function is given.
Here, we show how to derive this result and extend it to correlation functions of arbitrary order.
We start from the expression \eqref{CTRW-model} for the characteristic function for a CTRW with an exponential waiting-time distribution.
By definition, this is the Fourier transform of the probability density of finding the random walker at position $x$ at time $t$, starting from $x = 0$ at $t = 0$.
For an exponential waiting-time distribution the CTRW is homogeneous in both time and space and thus $K(k,t)$ also is the Fourier transform of the conditional probability density,
\begin{align}
K(k,t_2-t_1) = \int dx \ e^{i k x} P(x_1 + x,t_2 \vert x_1,t_1) ,
\end{align}
i.~e.~the properties of the random walk are independent of its starting point in time and space and only relative displacements matter.
Then, we can write the $n$-point probability density as,
\begin{align}
P(x_n,t_n; x_{n-1}, t_{n-1}; &\ldots; x_1,t_1) = \frac{1}{(2\pi)^n} \int dk_n \int dk_{n-1} \ldots \int dk_1 \ e^{-i \big(k_n (x_n - x_{n-1}) + k_{n-1} (x_{n-1}-x_{n-2}) + \ldots + k_1 x_1 \big)} \nn
& \hspace{3cm} K(k_n,t_n-t_{n-1}) K(k_{n-1},t_{n-1}-t_{n-2}) \ldots K(k_1,t_1) \nn
&= \frac{1}{(2\pi)^n} \int dk_n \int dk_{n-1} \ldots \int dk_1 \ e^{-i \big(k_n x_n + (k_{n-1} - k_n) x_{n-1} + (k_{n-2} - k_{n-1}) x_{n-1} + \ldots + (k_1-k_2) x_1 \big)} \nn
& \hspace{2cm} K(k_n,t_n-t_{n-1}) K(k_{n-1},t_{n-1}-t_{n-2}) \ldots K(k_1,t_1) .
\end{align}
The $n$-point correlation function is defined as,
\begin{align}
C_n(t_n,\ldots,t_1) = \int dx_n \int dx_{n-1} \ldots \int dx_1 \ x_n x_{n-1} \ldots x_1 P(x_n,t_n; x_{n-1}, t_{n-1}; \ldots; x_1,t_1).
\end{align}
By the properties of the Fourier transform, we have,
\begin{align}
\frac{1}{2\pi}\int dx_1 \int dk_1 \ x_1 e^{i (k_1-k_2) x_1} f(k_1) &= - \frac{i}{2\pi} \int dx_1 \int dk_1 \ e^{i (k_1-k_2) x_1} \partial_{k_1} f(k_1) \nn
&= -i \int d{k_1} \ \delta(k_1-k_2) \partial_{k_1} f(k_1) = - i \partial_{k_2} f(k_2).
\end{align}
Repeating this procedure, we accordingly find,
\begin{align}
C_n(t_n,\ldots,t_1) = (-i)^n \partial_{k} \bigg[ K(k,t_n-t_{n-1}) \partial_k \Big[ K(k,t_{n-1}-t_{n-2}) \partial_k \big[ \ldots \partial_k K(k,t_1) \big] \Big] \bigg]_{k=0} .
\end{align}
If $t_n = t_{n-1} = \ldots t_1 = t$, then, since we have $K(k,0) = 1$, this reduces to the familiar formula for the $n$-th moment,
\begin{align}
\av{x^n(t)} = (-i)^n \partial_k^n K(k,t) \Big\vert_{k = 0} .
\end{align}
For the explicit form Eq.~\eqref{CTRW-model}, we can express the $n$-point correlation function as a polynomial of the moments of the jump length distribution, e.~g.~for $n=4$,
\begin{align}
C_4(t_4,t_3,t_2,t_1) &=  \frac{t_1}{\tau_0} \av{\xi^4} + \frac{(t_4 + t_3 + 2 t_1) t_1}{\tau_0^2} \av{\xi^3} \av{\xi} + \frac{(t_3 + 2 t_2) t_1}{\tau_0^2} \av{\xi^2}^2 \nn
& \hspace{1cm} + \frac{(t_4 t_3 + 2 t_4 t_2 + 3 t_3 t_2) t_1}{\tau_0^3} \av{\xi^2} \av{\xi}^2 + \frac{t_4 t_3 t_2 t_1}{\tau_0^4} \av{\xi}^4  \nn
&=  \frac{t_1}{\tau_0} \av{\xi^4} + \frac{(t_3 + 2 t_2) t_1}{\tau_0^2} \av{\xi^2}^2,
\end{align}
where the latter expression is valid for a symmetric jump length distribution.

\section{Comparison with numerical simulations}  \label{app-num}
\begin{figure}[t]
\includegraphics[width=0.485\textwidth]{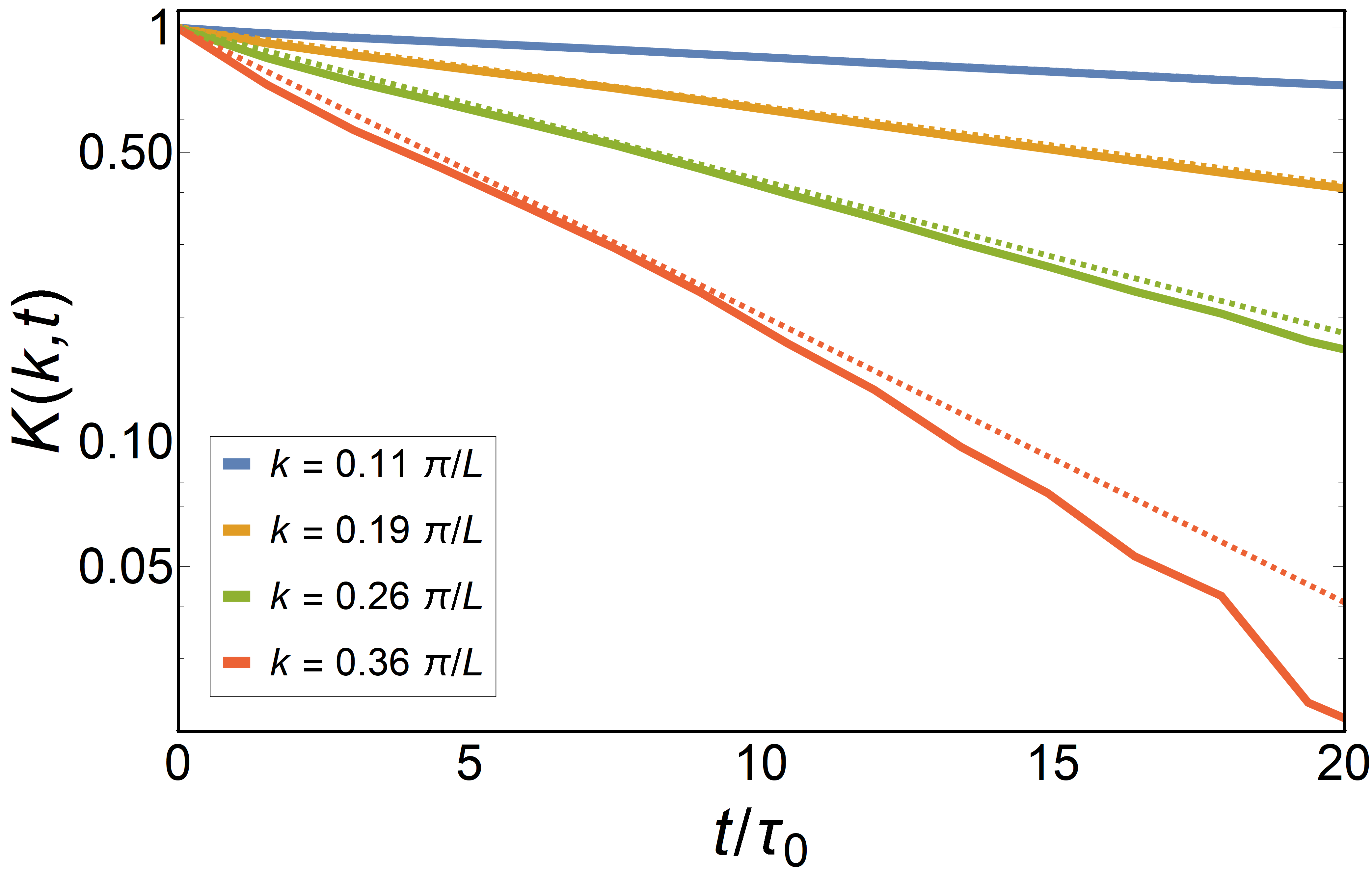}
\includegraphics[width=0.475\textwidth]{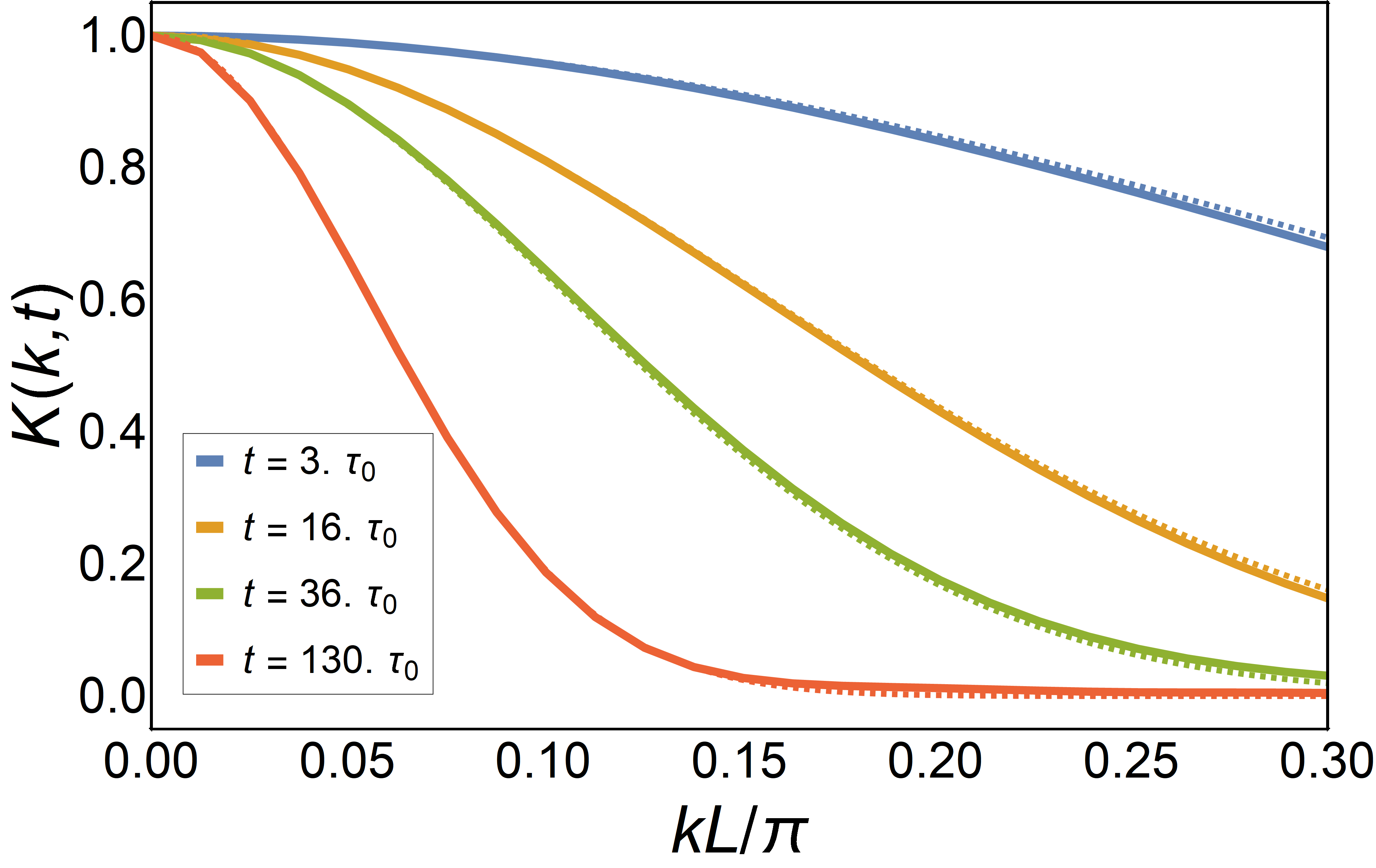}
\caption{The characteristic function Eq.~\eqref{CTRW-model} as a function of time (top) and wave vector (bottom) in the strong damping regime. 
The solid lines are obtained from numerical simulations of Eq.~\eqref{langevin} with a cosine potential and $U_0 = 5 \kb T$, the dashed lines correspond to the analytic result obtained from Eq.~\eqref{CTRW-model}. 
The remaining parameters of the numerical simulations are $m \gamma = 1$, $U_0 = 1$ and $L = 0.5$. \label{fig-charfunc-num-od}}
\end{figure}

\begin{figure}[t]
\includegraphics[width=0.485\textwidth]{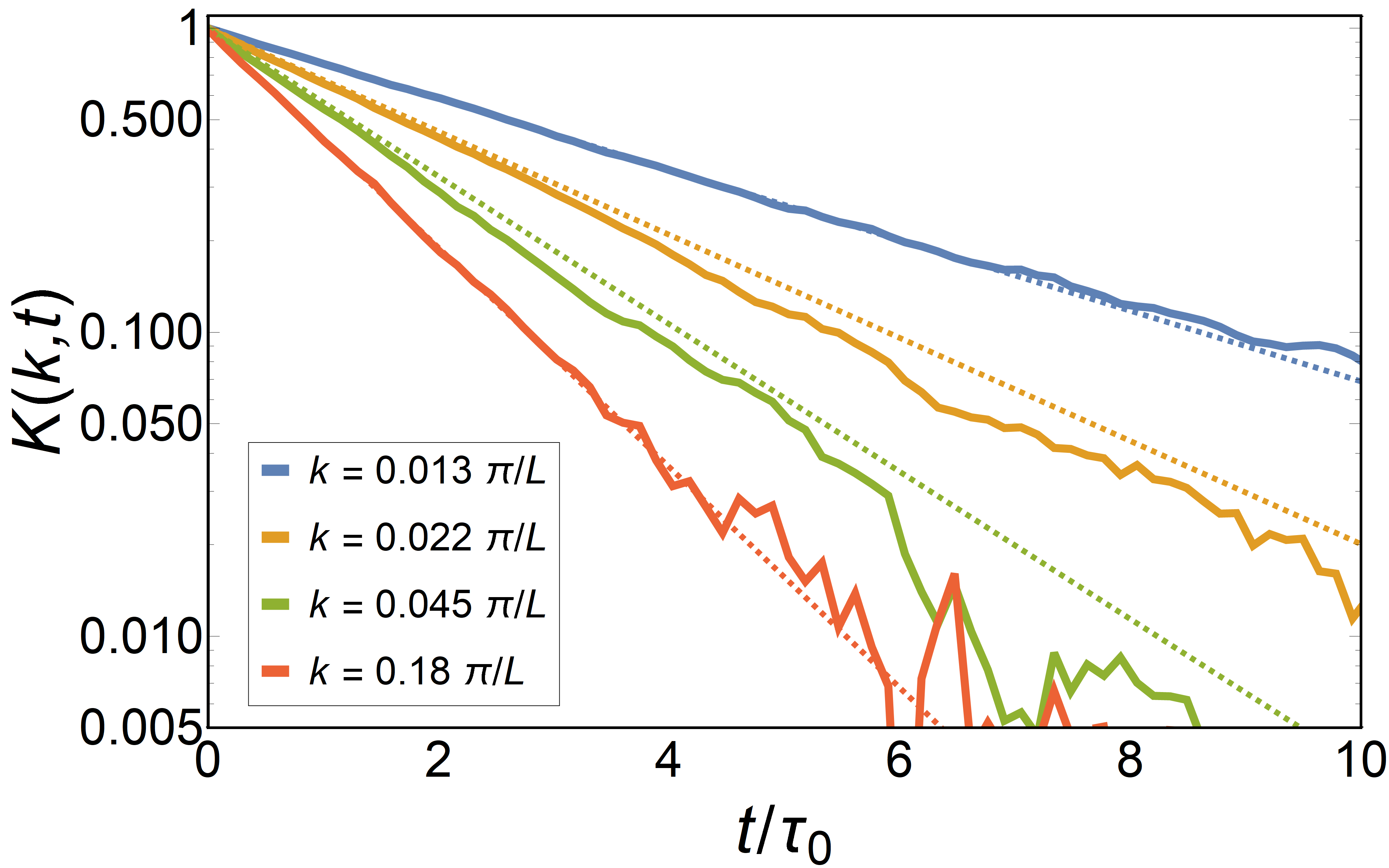}
\includegraphics[width=0.475\textwidth]{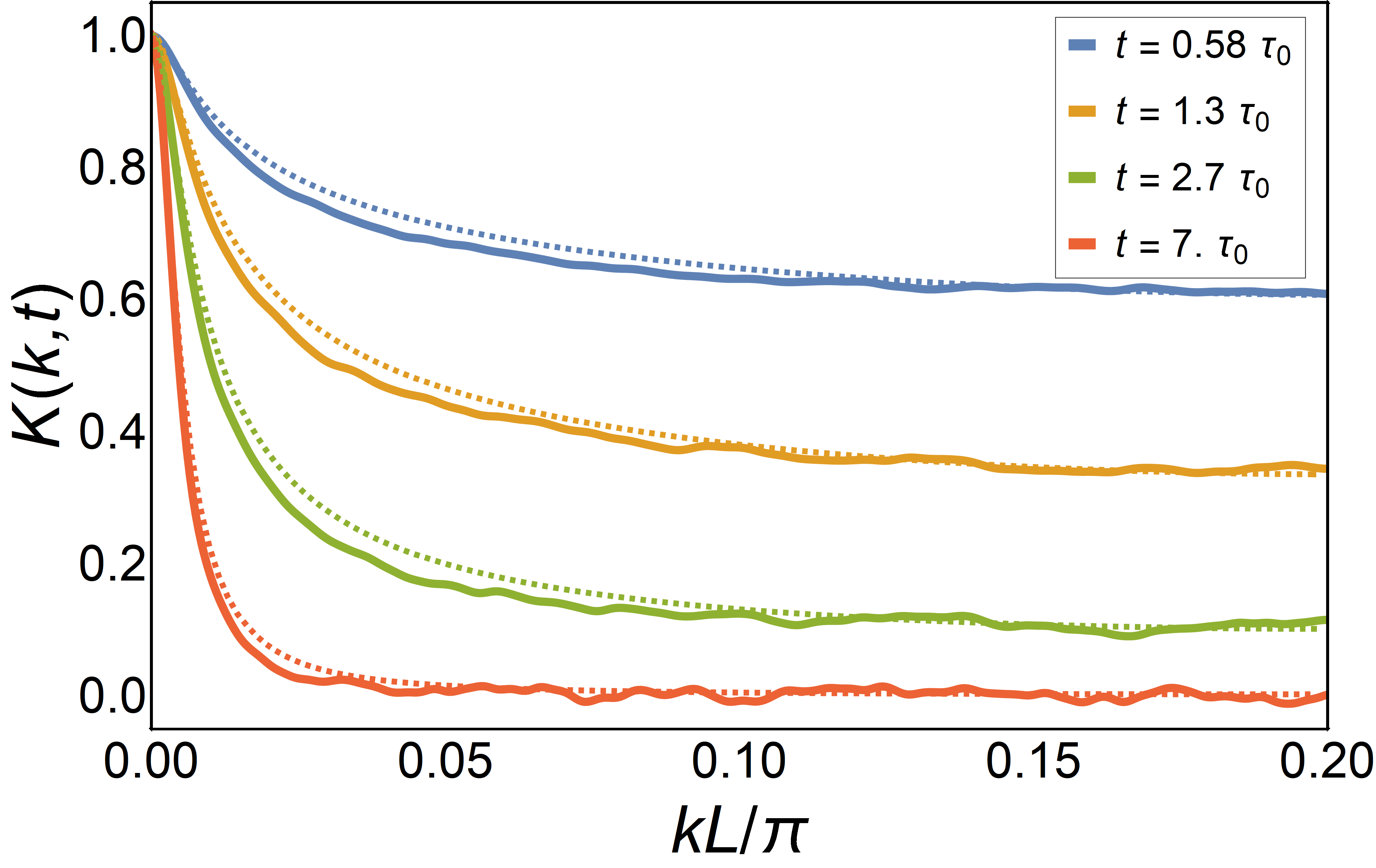}
\caption{The characteristic function Eq.~\eqref{CTRW-model} as a function of time (top) and wave vector (bottom) in the weak damping regime. 
The solid lines are obtained from numerical simulations of Eq.~\eqref{langevin} with the potential Eq.~\eqref{potential} and $U_0 = 5 \kb T$ and $\gamma = 0.01$ (corresponding to $\mathcal{E} = 0.023 \ \kb T$), the dashed lines correspond to the analytic result obtained from Eq.~\eqref{CTRW-model}. 
The remaining parameters of the numerical simulations are $m = 1$, $U_0 = 1$ and $L = 0.5$. \label{fig-charfunc-num}}
\end{figure}

In order to asses the validity of the CTRW model Eq.~\eqref{CTRW-model}, we compare its predictions to numerical simulations of the Langevin dynamics Eq.~\eqref{langevin}, both in the underdamped and overdamped  regimes.
In the numerical simulations, the microscopic escape time and the jump length distributions $\psi(\tau)$ and $\phi(\xi)$ can be obtained directly and exhibit excellent agreement with the theoretical predictions (see Figs.~\ref{fig-esc-jump-over} and \ref{fig-esc-jump-under}). 
However, directly observing these distributions is generally not possible for experimental data, where individual jump events may not be resolved due to limited spatial or temporal resolution. 
We therefore focus on comparing the theoretical results with the characteristic function, which can also be observed in the experiment. 
A central assumption entering the CTRW model is the exponential waiting time distribution, which is directly reflected in the exponential decay of the characteristic function Eq.~\eqref{CTRW-model}.
As shown in the left panel of Figs.~\ref{fig-charfunc-num-od} and \ref{fig-charfunc-num} for strong and weak damping, respectively, we indeed find that the characteristic function, obtained by Fourier-transforming the probability distribution observed in the simulations, decays exponentially with time.
As a consequence, we expect the qualitative diffusion process to be well-described by Eq.~\eqref{CTRW-model}

\begin{figure}[t]
\includegraphics[width=0.49\textwidth]{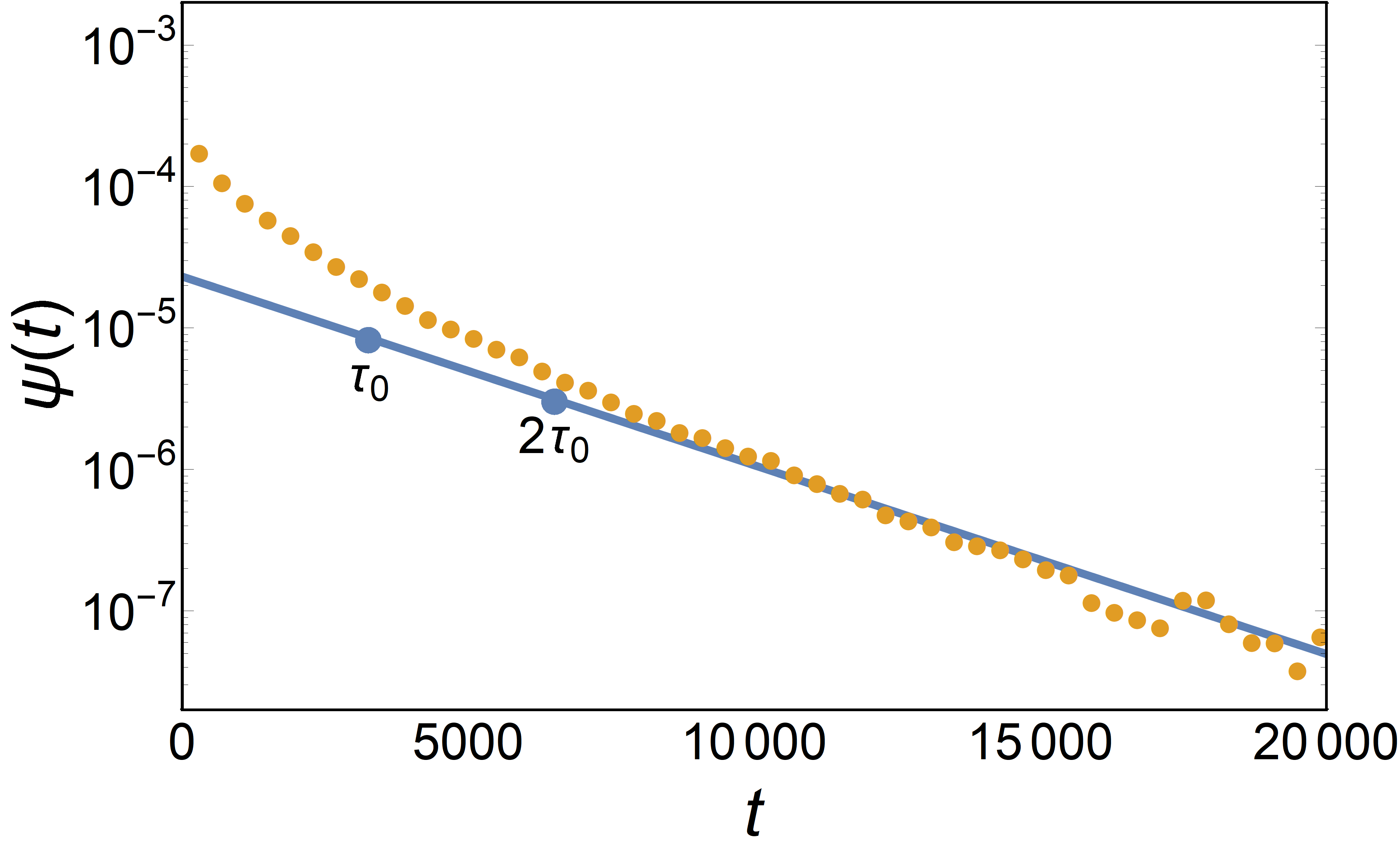}
\caption{Escape time distribution including repeated jump events. The orange dots are the distribution obtained from the numerical simulation, the blue line represents an exponential decay with time constant $\tau_0$ given by Eq.~\eqref{fpt-under-average}. Note that, compared to a purely exponential distribution, which is realized for $t \gtrsim 2 \tau_0$, the probability of short escape times is significantly enhanced, reflecting the finite thermalization time.} \label{fig-esc-jump-under-2}
\end{figure}
In the strong damping regime, we find almost perfect agreement between the analytical predictions and the simulation data, see Fig.~\ref{fig-charfunc-num-od}.
In the weak damping regime, the overall quantitative agreement between the predictions and the simulation results remains very good, with a slight discrepancy at intermediate values of $k$.
We suspect that this discrepancy is due to an effect not captured by the CTRW model: 
At the end of a jump, the particle does not immediately thermalize in the potential well; instead, its energy remains close to the barrier energy for a time on  the order of the equilibration time $\tau_\text{eq} = 1/\gamma$.
As a consequence the probability of making another jump is increased.
Note that the left panel of Fig.~\ref{fig-esc-jump-under} represents the escape time distribution of particles whose initial energy distribution is given by the stationary distribution Eq.~\eqref{energy-dist-stat}, i.~e.~which are initially thermalized in a potential well and thus does not include repeated jump events.
On the other hand, the escape time distribution of all particles (thermalized or not) shows a marked enhancement at short times as a consequence of repeated jump events, see Fig.~\ref{fig-esc-jump-under-2}.
We expect this effect to impact the characteristic function at wave vectors corresponding to length scales of a few single jumps, which for the parameters in Fig.~\ref{fig-charfunc-num} is $k L/\pi \approx 0.03$. 
This is precisely the range in which we observe the discrepancy between the simulation results and the CTRW model in Fig.~\ref{fig-charfunc-num}.
Note that, since the thermalization is almost instantaneous in the strong damping regime, this effect does not appear in Fig.~\ref{fig-charfunc-num-od}. 
\end{widetext}

\end{document}